\documentclass[runningheads]{llncs}
\usepackage{comment}
\usepackage{graphicx}
\usepackage{multirow}
\usepackage{amsmath}
\usepackage{subfigure}
\usepackage{makecell}
\begin{document}

\title{TST: Time-Sparse Transducer for Automatic Speech Recognition}

% INITIAL SUBMISSION 
\def\CICAISubNumber{53}  % Insert your submission number here
%\begin{comment}
\titlerunning{TST: Time-Sparse Transducer for Automatic Speech Recognition} 
\authorrunning{Xiaohui Zhang et al.} 
\author{Xiaohui Zhang\inst{1,2} 
\and
Mangui Liang\inst{1} 
\and
Zhengkun Tian\inst{2}
\and
Jiangyan Yi\inst{2}
\and
Jianhua Tao\inst{3}
}
% \institute{Paper ID \CICAISubNumber}
\institute{School of Computer and Information Technology, Beijing Jiaotong University, Beijing, China
% \email{\{21120320,mgliang\}@bjtu.edu.cn}\\
\and
Institute of Automation, Chinese Academy of Science, Beijing, China
% \email{\{zhengkun.tian,jiangyan.yi\}@nlpr.ia.ac.cn}\\
\and
Department of Automation, Tsinghua University, Beijing, China
% \email{jhtao@tsinghua.edu.cn}}
}
%\end{comment}
%******************

% CAMERA READY SUBMISSION
\begin{comment}
% \titlerunning{Abbreviated paper title}
% If the paper title is too long for the running head, you can set
% an abbreviated paper title here
%
%
\authorrunning{F. Author et al.}
% First names are abbreviated in the running head.
% If there are more than two authors, 'et al.' is used.
\end{comment}
%******************
\maketitle              % typeset the header of the contribution

\begin{abstract}
End-to-end model, especially Recurrent Neural Network Transducer (RNN-T), has achieved great success in speech recognition. However, transducer requires a great memory footprint and computing time when processing a long decoding sequence. To solve this problem, we propose a model named time-sparse transducer, which introduces a time-sparse mechanism into transducer. In this mechanism, we obtain the intermediate representations by reducing the time resolution of the hidden states. Then the weighted average algorithm is used to combine these representations into sparse hidden states followed by the decoder. All the experiments are conducted on a Mandarin dataset AISHELL-1. Compared with RNN-T, the character error rate of the time-sparse transducer is close to RNN-T and the real-time factor is 50.00\% of the original. By adjusting the time resolution, the time-sparse transducer can also reduce the real-time factor to 16.54\% of the original at the expense of a 4.94\% loss of precision.

\end{abstract}

\section{Introduction}
In recent years, significant advancements have been made in end-to-end speech recognition models, including the connectionist temporal classification (CTC) \cite{graves2006connectionist,graves2014towards,amodei2016deep}, attention-based sequence-to-sequence models (AED) \cite{bahdanau2014neural,vaswani2017attention,dong2018speech,kim2017joint}, and recurrent neural network transducer (RNN-T) \cite{graves2012sequence,graves2013speech,rao2017exploring,he2019streaming,Han2020ContextNetIC,9053896}. The CTC algorithm, employed by many models, performs frame-level decoding by converting speech sequences to corresponding label sequences. However, this method relies on the assumption of conditional independence among speech frames, making it unable to effectively model the dependencies between outputs. On the other hand, the RNN-T leverages its recurrent structure to overcome the conditional independence assumption and optimizes acoustic and language components jointly through the introduction of language and joint networks. Consequently, the RNN-T has found success in online automatic speech recognition (ASR) systems \cite{kannan2019large,9053600}.\par
Despite its advantages, the RNN-T imposes a higher memory demand compared to AED and CTC methods \cite{rao2017exploring,li2019improving}. During the forward-backward pass of the RNN-T, posteriors are calculated at each point within the grid composed of the encoder and prediction network. Computing a long decoding sequence in this grid consumes more memory and time than the aforementioned methods, making the vocabulary less dependent on training/inference speech and more reliant on sequence length \cite{9414502,6854533,9413854}. Therefore, reducing memory consumption and improving computing speed are crucial for deploying RNN-T models on low-resource devices \cite{9413535,9003906,9443456}.\par
This paper proposes a model, named the time-sparse transducer (TST), designed to address the memory cost and computing time consumption of the RNN-T. Our approach consists of a convolutional front end, an acoustic encoder, a time-sparse mechanism, and a decoder. The encoder maps input acoustic frames into high-level representations, while the decoder, analogous to a conventional language model, combines these representations to produce a distribution over the output target through a softmax layer. The prediction network and joint network collectively form the decoder. The time-sparse mechanism reduces the time resolution by decomposing the encoder's hidden states into intermediate encoded representations using a sliding pooling window. These representations are then combined into sparse hidden states using a weighted average algorithm, and subsequently fed to the joint network. As a result, the sequence length of the sparse hidden states outputted by the time-sparse mechanism is significantly smaller than that of the encoder. This compression in the length of hidden states effectively reduces the GPU memory footprint and computing time. Additionally, introducing an attention mechanism \cite{vaswani2017attention} during the combination of intermediate encoded representations enhances the coefficients of representations with valuable information and suppresses noisy representations. Furthermore, the attention coefficients contribute to a lower character error rate (CER) for our model compared to the RNN-T baseline, as demonstrated through experiments conducted on the AISHELL-1 dataset.\par
The remaining sections of this paper are organized as follows: Section \ref{background} provides an overview of the RNN-T method, while Section \ref{Time-Sparse Transducer} describes the structure of the time-sparse mechanism and highlights key considerations during the generation of sparse representations. This section also presents the strategies for generating the weighted average coefficients. In Section \ref{Experiment}, we detail the experiments conducted and their respective results. Finally, in Section \ref{conclusion}, we present our conclusions, summarizing the key aspects and effects of the TST model.

\section{Background}
\label{background}
Our proposed approach is based on the Recurrent Neural Network Transducer (RNN-T) model. In this section, we provide an overview of the RNN-T structure, training strategy, and decoding process \cite{tian2022hybrid}. \par
% \cite{tian2022hybrid,tian2019self,sak2015learning}.\par
The RNN-T consists of two distinct networks: the acoustic encoder and the prediction network, which are connected through the joint network. The acoustic encoder maps an input frame $\textbf{x}_t$ to a hidden state vector $\textbf{h}_t$. The linguistic state vector $\textbf{g}_u$ is generated by appending the prediction "non-blank" symbol from the previous time step to the prediction network. The joint network is a feed-forward network that combines the hidden state vector $\textbf{h}_t$ and the linguistic state vector $\textbf{g}_u$ as follows:
%\begin{equation}
%\setlength{\abovedisplayskip}{3pt}
%m_{t,u}=JoinNet(h_t,g_u)
%\label{eq_1}
%\setlength{\belowdisplayskip}{3pt}
%\end{equation}
\begin{equation}
\setlength{\abovedisplayskip}{3pt}
\textbf{z}_{t,u}=f_{Activate}(W\textbf{h}_{t}+V\textbf{g}_{u}+b)
\setlength{\belowdisplayskip}{3pt}
\label{eq_2}
\end{equation}
Here, $W$ and $V$ are weight matrices, $b$ is a bias vector, and $f_{Activate}$ represents an activation function such as Tanh or ReLU. The output $\textbf{z}_{t,u}$ is then linearly transformed:
\begin{equation}
\setlength{\abovedisplayskip}{6pt}
\textbf{m}_{t,u}={\rm{Linear}}(\textbf{z}_{t,u})
\label{eq_4}
\setlength{\belowdisplayskip}{6pt}
\end{equation}
To obtain the posterior probability distribution $p(k|t,u)$ over the next output symbol, a softmax function is applied to $\textbf{m}_{t,u}$:
\begin{equation}
\setlength{\abovedisplayskip}{6pt}
p(k|t,u)=\rm{Softmax}(\textbf{m}_{t,u})
\label{eq_3}
\setlength{\belowdisplayskip}{6pt}
\end{equation}
% The probability distribution needs to be computed at each point in the grid composed of the prediction network and acoustic encoder. Then RNN-T sums the probabilities of all possible paths by the forward-backward algorithm. For this reason, The RNN-T processing is very memory demanding \cite{rao2017exploring}. The loss function of RNN-T is the negative log-likelihood of the target sequence $y*$.
The probability distribution is computed at each point in the grid formed by the prediction network and the acoustic encoder. The RNN-T model employs the forward-backward algorithm to sum the probabilities of all possible paths. However, this processing approach results in significant memory requirements \cite{rao2017exploring}. The loss function for RNN-T is defined as the negative log-likelihood of the target sequence $y*$:
\begin{equation}
\setlength{\abovedisplayskip}{6pt}
\mathcal{L}_{RNN-T}=-{\rm{ln}}(P(y*|x))
\label{eq_5}
\setlength{\belowdisplayskip}{6pt}
\end{equation}
In terms of inference, the RNN-T performs frame-by-frame computation, which can be slow when processing long sequences. The decoder employs beam search and greedy search methods to identify the most likely sequence as the output of the network \cite{graves2012sequence}.\par
Overall, the RNN-T model exhibits a distinctive architecture, comprising an acoustic encoder, prediction network, and joint network. The forward-backward algorithm is employed to compute probabilities, while the negative log-likelihood serves as the loss function. The inference process can be time-consuming for long sequences, and decoding methods like beam search and greedy search are employed to obtain the output sequence.
\begin{figure}[t]
%	\figure[The Architecture of SST]{
\centerline{\includegraphics[width=0.6\textwidth]{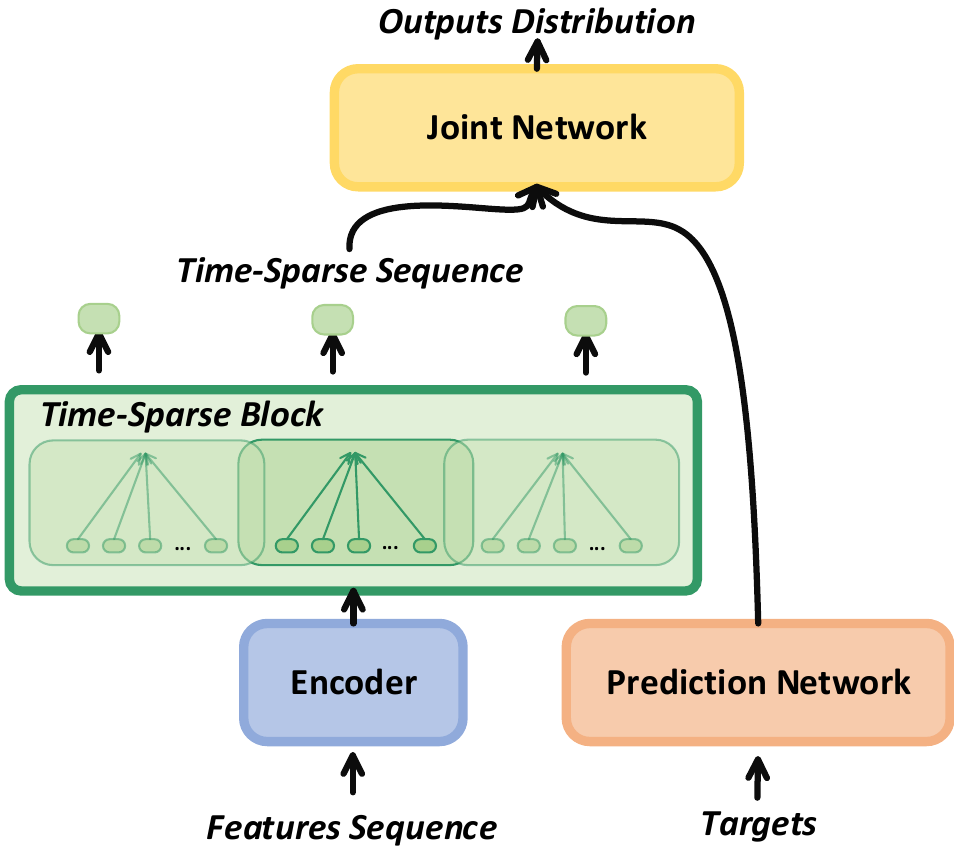}}
\caption{Illustration of the structure of our proposed model time-sparse transducer. Compared with the RNN-T, a sparse block in time is introduced between the encoder and the joint network as shown in the figure. All intermediate representations are generated by sliding a window, which is shown in the Time-Sparse Block of this figure, on hidden states with two hyperparameters: window length and stride. Both of them can be set before the training process and would make observed disparity on the reasoning output. After that, the time-sparse block combines the encoded representations through the weighted average algorithm and feeds them into the joint network. The sequence length of the sparse hidden state is reduced after this process.}
\label{fig.1}
\end{figure}
\section{Methodology}
\label{Time-Sparse Transducer}
Our proposed method, Time-Sparse Transformer (TST), is based on the decomposition of encoded hidden states using a time-sparse mechanism. The time-sparse mechanism consists of two components: a decomposition of hidden states based on a sliding pooling window and a combination process of sparse hidden states based on the weighted average algorithm. In this section, we provide detailed explanations of each component.
\subsection{The decomposition based on sliding pooling window}
We first reduce the time resolution by decomposing the encoded hidden states into intermediate encoded representations by the TST algorithm. In the time-sparse mechanism, the intermediate representations are generated by sliding a window on hidden states with fixed window length and stride. Through this process, the information carried by the output of the encoder can be spread over various encoded representations $\mathbf{r_k}$ with smaller time resolution than hidden states as Equation \ref{win_f}:
\begin{equation}
\setlength{\abovedisplayskip}{6pt}
[\mathbf{r_1},\mathbf{r_2},\mathbf{r_3},...\mathbf{r_k},...\mathbf{r_n}]=f_{win}(\mathbf{x}, input, length, stride)
\label{win_f}
\setlength{\belowdisplayskip}{6pt}
\end{equation}
where the $\mathbf{x}$ and $\mathbf{r_k}$ are the input and intermediate encoded representation output by decomposition respectively and $n$ is the number of intermediate representations. The window can overlap partly during sliding to ensure continuous information between each representation. The pooling process is primarily affected by two factors, namely, the window length affecting the size of the sliding window, and the window stride affecting the size of the overlap between windows. The time resolution of the intermediate encoded representation decreases as the window length and stride size increase. So a large window length and stride size will cause the loss of detailed information but retain more global information when the encoder output is decomposed. After the decomposition, the intermediate encoded representations can be calculated with less computational effort in a shorter time due to the smaller sequence length than encoded hidden states. Conversely, setting a smaller window length and stride size can retain more detailed information, but also generate representations with greater sequence length than setting smaller sequence length and size, thus increasing the processing time and memory requirements of the time-sparse mechanism.\par
\subsection{The combination based on weighted average algorithm}
After decomposing the encoded hidden states by sliding window, we combine the encoded representations through the weighted average algorithm and feed them into the joint network.
In this process, the weighted average algorithm does not need to change the sequence length of the input and output, so that the time resolution of the sparse hidden state fed into the joint network by TST is much smaller than that of the hidden state fed into the joint network by RNN-T.
\subsubsection{Combine with absolute average}
In weighted average combination, an intuitive way is that all weight coefficients are initialized to be identical, as shown in Equation \ref{eq_8}.
\begin{equation}
\setlength{\abovedisplayskip}{6pt}
\textbf{h}^{'}_{t}=\sum_{k=1}^{n}\frac{1}{n}\cdot\textbf{r}_k%(r_1,r_2,r_3...r_n)
\label{eq_8}
\setlength{\belowdisplayskip}{6pt}
\end{equation}
% where $\textbf{r}$ is the intermediate encoded representation output by decomposition and $n$ is the number of intermediate representations. 
This method reduces the time it consumes to calculate the coefficient but ignores the difference in the information carried by intermediate encoded representations. 
\subsubsection{Combine with learnable coefficients}
In addition to the absolute average, the weight coefficient can also be initialized with a set of random coefficients and jointly optimized with other parameters through the RNN-T loss function during model training \cite{ostmeyer2019machine}. This process is illuminated as 
%\begin{equation}
%\setlength{\abovedisplayskip}{6pt}
%W_{i+1}=Optim(W_i,\mathcal{L}_{RNN-T})
%\label{eq_12}
%\setlength{\belowdisplayskip}{6pt}
%\end{equation} 
\begin{equation}
\setlength{\abovedisplayskip}{6pt}
\textbf{h}^{'}_{t}=\sum_{k=1}^{n}w_k\cdot\textbf{r}_k%(w_1,w_2,w_3...w_n)\cdot(r_1,r_2,r_3...r_n)
\label{eq_9}
\setlength{\belowdisplayskip}{6pt}
\end{equation}
where $w_k$ is the learnable coefficient. The precision of the TST can be improved by increasing the coefficient of the intermediate encoded representation with the information that has a positive impact on the prediction and suppressing the coefficient of the representation that carries noise information. 
\subsubsection{Combine with attention mechanism}
Inspired by the self-attention mechanism \cite{vaswani2017attention}, we introduce the attention mechanism to calculate the weighted average coefficients. The attention weights $\alpha_k$ are computed as follows:
\begin{equation}
\setlength{\abovedisplayskip}{6pt}
{\rm{e}}_k=\rm{Linear}(\textbf{r})
\label{eq_10}
\setlength{\belowdisplayskip}{6pt}
\end{equation}
\begin{equation}
\setlength{\abovedisplayskip}{6pt}
\alpha_k={\rm{Softmax}(e}_k)
\label{eq_12}
\setlength{\belowdisplayskip}{6pt}
\end{equation}
\begin{equation}
\setlength{\abovedisplayskip}{6pt}
\textbf{h}^{'}_{t}=\sum_{k=1}^{n}\alpha_k\cdot\textbf{r}_k
\label{eq_11}
\setlength{\belowdisplayskip}{6pt}
\end{equation}
In this case, the attention mechanism allows the model to focus more on positive information. By calculating the attention weights, TST ensures that the intermediate encoded representations with significant contributions are given higher coefficients in the combination process.\par
In conclusion, the decomposition of the encoded hidden states and the combination of intermediate encoded representations occur only between the encoder and the joint network. From the decoder's perspective, the features inputted to the decoder in TST are indistinguishable from those in RNN-T. Therefore, any decoder used for RNN-T can be employed for TST.
\section{Experiments and Results}
\label{Experiment}
\subsection{Experimental setup}
All of our experiments are conducted on a public Mandarin speech corpus AISHELL-1 \cite{bu2017aishell}. We use 80-dimension Mel-filter Bank coefficients (FBank) features with 3-dim pitch features computed on 25 ms window with 10 ms shift, which is known to be effective in Mandarin speech recognition. The 4234 characters (including a padding symbol \texttt{<PAD>}, an unknown token \texttt{<UNK>}, a begin-of-sequence token \texttt{<BOS>} and an end-of-sequence token \texttt{<EOS>}) are chosen as modeling units.\par
For the baseline RNN-T model, the front-end convolutional block followed by the encoder consists of two 2D-Convolution layers with a ReLU activation, stride size 2, channels 384, kernel size 3, and output size 384. The acoustic encoder consists of 12 transformer blocks with 4 heads in multi-head attention. The feed-forward size of the encoder is 384 and the hidden size is 768. We utilize three types of decoders. The first is a transformer decoder with 6 blocks and 4 heads in multi-head attention \cite{moritz2020streaming,tian2020spike,dong2018speech}; The second is a state-less decoder \cite{ghodsi2020rnn}, and the third is an RNN decoder with 2-layer Long Short-Term Memory (LSTM) model. The configuration of TST is the same as RNN-T, except for introducing a time-sparse mechanism between the RNN-T encoder and joint network. Experiments are conducted on the sliding pooling window length from 1 to 10 with fixed window stride 1 and stride from 1 to 10 with window length 10. The generation strategies of intermediate encoded representation coefficients are absolute equality (AE), optimization of the random initialization coefficients through the RNN-T loss function (LC), and calculation through the self-attention mechanism (SA). We utilize the CER to evaluate the accuracy of different models and the real-time factor (RTF) to evaluate the inference speed.
\begin{figure}[t]
%	\figure[The Architecture of SST]{
\centerline{\includegraphics[width=0.6\textwidth]{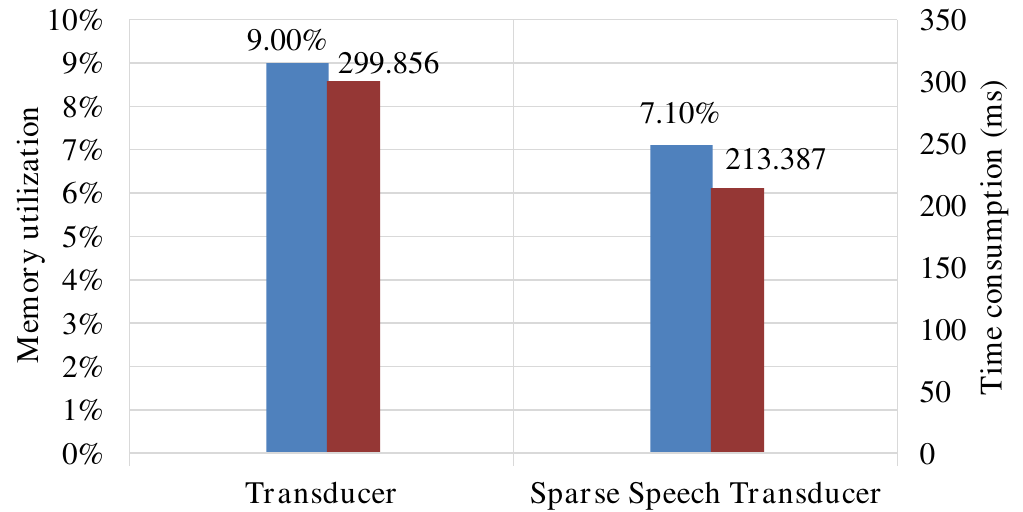}}
\caption{Illumination of the GPU memory utilization and time consumption of RNN-T and TST on batch data. Both experiments are conducted on a Tesla K80 GPU.}
\label{fig.2}
\end{figure}
\subsection{Results}
\subsubsection{The comparison of GPU consumption between RNN-T and TST}
%\begin{figure}
%	\subfigure[The comparison of GPU demanding]{
%		\centerline{\includegraphics[width=0.8\columnwidth]{GPU_comprison.pdf}}}
%	
%	\subfigure[The comparison of Time (ms) consumption]{
%		\centerline{\includegraphics[width=0.8\columnwidth]{Time_comprison.pdf}}}
%	\caption{(a) and (b) illuminate the GPU consumption and computing time of RNN-T and TST on a batch data. Both experiments are conducted on a Tesla K80 GPU.}
%	\label{fig.2}
%\end{figure}
We conducted a comparative analysis of GPU memory utilization and time consumption between RNN-T and TST, as depicted in Figure \ref{fig.2}. The primary objective was to evaluate the impact of the time-sparse mechanism on these performance metrics.
Our results revealed that the utilization of GPU memory decreased from 9\% to 7.1\% when utilizing TST instead of RNN-T. Additionally, the time consumption exhibited a noticeable improvement, reducing from 299.856 ms to 213.387 ms when processing batch data. These findings clearly demonstrate that TST achieves lower GPU occupancy and faster computation speed compared to RNN-T.
By incorporating the time-sparse mechanism into RNN-T, we effectively reduce the sequence length of the hidden states received by the prediction network. Consequently, this reduction in sequence length leads to decreased GPU occupancy and significantly reduces the time required for subsequent computations.
\begin{table}[t]
\centering
\caption{Comparison of our TST with the RNN-T after the downsampling process along time. To illustrate the disparity between before and after time sparing, we also show the performance of the commonly used RNN-T as the baseline. The time resolutions of both TST and downsampling RNN-T are reduced to 1/10 that of the RNN-T.}
\label{table3}
\small
\setlength{\tabcolsep}{3.0pt}
\begin{tabular}{|c|c|c|c|c|c|c|}
\hline
\multirow{2}*{Model}& \multirow{2}*{RNN-T}& \multirow{2}*{Downsampling}& \multicolumn{3}{|c|}{TST} \\
\cline{4-6}
~ & ~& ~& AE & LC & SA\par\\
\hline
\hline
CER (\%) & 7.824 & 43.926 & 15.784 & 13.411 & 12.760 \par\\
\hline
RTF		 & 0.122 & 0.019 & 0.021 & 0.021 & 0.021\par\\
\hline
\end{tabular}
\label{tab3}
\end{table}
\subsubsection{The influence of downsampling in the encoded hidden state of TST and RNN-T}
In this section, we evaluate the performance of TST and RNN-T when the time resolution is reduced. Table \ref{tab3} presents the results, highlighting that there is no significant difference in terms of Real-Time Factor (RTF) between TST and RNN-T. Both models achieve a similar RTF reduction, approximately to 1/6 of the original value, when the time resolution is decreased to 1/10. However, it is important to note that downsampling the data along the time axis leads to a substantial loss in accuracy, amounting to 36.102\%.\par
Regarding our TST, it incorporates a time-sparse mechanism, enhanced by a self-attention block, which effectively mitigates the accuracy loss. Specifically, TST achieves a remarkable reduction in accuracy loss to 4.936\%. The introduction of the time-sparse mechanism, with its self-attention component, enables TST to better preserve the relevant information in the sparse representations, thus significantly minimizing the adverse effects of downsampling on accuracy.
\begin{table}[t]
\centering
\caption{Comparison of TST with different sliding pooling window lengths. The window stride of all experiments is 1.}
% \resizebox{!}{1.25cm}{
\setlength{\tabcolsep}{1.5pt}
\begin{tabular}{|c|c|c|c|c|c|c|}
\hline
\multirow{2}*{Window Length}& \multicolumn{2}{|c|}{AE} & \multicolumn{2}{|c|}{LC} & \multicolumn{2}{|c|}{SA} \par\\
\cline{2-7}
~ & CER(\%) & RTF & CER(\%) & RTF & CER(\%) & RTF\par\\
\hline
\hline
10 & 9.454 & 0.160 & 9.213 & 0.161 & 8.714 & 0.158 \par\\
\hline
8  & 9.225 & 0.162 & 8.521 & 0.163 & 8.231 & 0.159 \par\\
\hline
6  & 9.013 & 0.160 & 8.244 & 0.163 & 7.942 & 0.160 \par\\
\hline
4  & 8.562 & 0.161 & 7.685 & 0.162 & \textbf{7.418} & 0.161 \par\\
\hline
2  & 8.259 & 0.162 & 7.812 & 0.164 & 7.533 & 0.159 \par\\
\hline
1  & 8.205 & 0.162 & 8.139 & 0.163 & \textbf{7.824} & 0.159 \par\\
\hline
\end{tabular}
% }
\label{tab1}
\end{table}
\begin{table}[t]
\centering
\caption{Comparison of TST with different sliding pooling window strides. The window length of all experiments is 10.}
% \resizebox{!}{1.24cm}{
\setlength{\tabcolsep}{1.5pt}
\begin{tabular}{|c|c|c|c|c|c|c|}
\hline
\multirow{2}*{Window Stride}& \multicolumn{2}{|c|}{AE} & \multicolumn{2}{|c|}{LC} & \multicolumn{2}{|c|}{SA} \par\\
\cline{2-7}
~ & CER(\%) & RTF & CER(\%) & RTF & CER(\%) & RTF\par\\
\hline
\hline
10 & 15.784 & 0.021 & 13.411 & 0.022 & 12.760 & \textbf{0.021}\par\\
\hline
8  & 13.598 & 0.039 & 11.531 & 0.039 & 11.320 & 0.039\par\\
\hline
6  & 11.347 & 0.050 & 10.338 & 0.050 & 9.672  & 0.049\par\\
\hline
4  & 10.442 & 0.061 & 9.866  & 0.061 & 8.831  & 0.060\par\\
\hline
2  & 10.215 & 0.093 & 9.734  & 0.093 & 8.799  & 0.093\par\\
\hline
1  & 9.454  & 0.160 & 9.213  & 0.161 & 8.714  & 0.158\par\\
\hline
\end{tabular}
% }
\label{tab2}
\end{table}

\subsubsection{The influence of the sliding pooling window length and stride on the model performance}
This section presents a comparative analysis of models employing different sliding pooling window lengths and stride sizes. The experimental results, as depicted in Tables \ref{tab1} and \ref{tab2}, clearly demonstrate that reducing the window length and stride size improves the accuracy of the models. When a large window with a large step size slides over the encoded hidden state, it leads to the loss of more detailed information. Conversely, a small window with a small step size preserves more information but results in a longer sequence length. The outcomes reveal that, for TST, the weighted average combination approach effectively reduces the CER when the window length is less than 4, accomplishing this by increasing the weight assigned to positive information while suppressing the weight assigned to noise. Moreover, employing the SA yields the lowest CER among the tested approaches. However, it should be noted that an increase in window length diminishes the information content, thereby decreasing the accuracy of TST. Notably, the tables demonstrate that the RTF remains largely unaffected by changes in window length but decreases as the stride size increases. This observation highlights that the inference speed of TST primarily depends on the decoding sequence length rather than the scale of intermediate encoded representation.
\subsubsection{The experimental results of different window types}
This section primarily investigates the impact of different strategies employed for generating weighted average coefficients within the Time-Sparse Transformer (TST) framework. Notably, these strategies have no bearing on the decoding sequence length, thereby ensuring consistent computational speed for TST. The experimental findings presented in Table \ref{tab3}, \ref{tab1}, and \ref{tab2} provide compelling evidence regarding the efficacy of the SA strategy, which yields the lowest CER on the AISHELL-1 dataset. Comparing the LC approach with the SA strategy, it becomes evident that the inclusion of the attention mechanism significantly aids the model in learning the weighted average coefficients. Specifically, the self-attention mechanism strengthens the coefficient associated with intermediate encoded representations that convey pertinent information while simultaneously diminishing the coefficient assigned to representations containing noise. Consequently, the resulting sparse hidden state exhibits enhanced recognition capabilities pertinent to the target task. These results shed light on the discriminative power and adaptability provided by the self-attention mechanism within the TST framework, reinforcing its effectiveness for optimizing speech recognition performance.
\begin{table}[t]
\centering
\caption{The Result of RNN-T and TST with different decoders. Both window length and window stride are 4. The window type is SA.}
% \label{table4}
\small
\setlength{\tabcolsep}{22pt}
\begin{tabular}{|c|c|c|}
\hline
Model & CER(\%) & RTF\par\\
\hline
\hline
RNNT-T    & 7.979 & 0.508\par\\
\hline
TST-T (Ours)   & \textbf{7.744} & \textbf{0.285}\par\\
% \hline
\hline
RNNT-S    & 7.824 & 0.122\par\\
\hline
TST-S (Ours)   & \textbf{7.528} & \textbf{0.061}\par\\
% \hline
\hline
RNNT-R   & 11.454 & 0.397\par\\
\hline
TST-R (Ours)   & \textbf{10.193} & \textbf{0.222}\par\\
\hline
\end{tabular}
\label{tab4}
\end{table}
\begin{table}[t]
\centering
\caption{The results of RNN-T and TST with different decoders. Both window length and window stride are 10. The window type is SA.}
% \label{table5}
\small
\setlength{\tabcolsep}{22pt}
\begin{tabular}{|c|c|c|}
\hline
Model & CER(\%) & RTF\par\\
\hline
\hline
RNNT-T    & 7.979 & 0.508\par\\
\hline
TST-T (Ours)   & 12.934 & \textbf{0.084}\par\\
% \hline
\hline
RNNT-S    & 7.824 & 0.122\par\\
\hline
TST-S (Ours)   & 12.760 & \textbf{0.021}\par\\
% \hline
\hline
RNNT-R   & 11.454 & 0.397\par\\
\hline
TST-R (Ours)   & 15.968 & \textbf{0.089}\par\\
\hline
\end{tabular}
\label{tab5}
\end{table}
\subsubsection{The experimental results of sliding pooling window with different decoders}
This section focuses on comparing the performance of TST and RNN-T with different decoders, namely RNN-T with Transformer decoder (RNNT-T), RNN-T with State-Less decoder (RNNT-S), RNN-T with RNN decoder (RNNT-R), TST with Transformer decoder (TST-T), TST with State-Less decoder (TST-S), and TST with RNN decoder (TST-R). Experiments were conducted to evaluate their performance in terms of CER and recognition speed.\par
Table \ref{tab4} presents the results, indicating that TST consistently achieves a lower CER and higher recognition speed compared to RNN-T in all three types of decoder. Specifically, when comparing TST-S with RNNT-S, it was found that the CER of TST-S is very close to that of RNN-T, while the RTF is reduced to 50\% of RNNT-S. These findings highlight that TST-S achieves comparable accuracy to RNN-T while significantly improving computational efficiency.\par
Furthermore, Table \ref{tab5} provides insights when the time resolution is reduced to 1/10. The CERs of models with different decoders are observed to increase, but the RTFs are reduced to 16.535\% to 22.418\% of the original values. Notably, among all the experiments, TST-S achieves the highest accuracy and its RTF is reduced to 17.213\% of that of RNNT-S. These results demonstrate that TST offers the potential to improve both recognition accuracy and computational efficiency across various decoder types. 
% The findings suggest that TST outperforms RNN-T in terms of CER and recognition speed across different decoder types. TST-S shows comparable accuracy to RNN-T while significantly reducing computational requirements. 
Moreover, when the time resolution is decreased (see Table \ref{tab4}), TST maintains superior performance in terms of accuracy and computational efficiency compared to RNN-T.
\section{Conclusion}
\label{conclusion}
In this study, we have introduced a novel model called the time-sparse transducer, which incorporates a time-sparse mechanism between the recurrent neural network (RNN) transducer encoder and the prediction network. Our proposed model offers several advantages compared to conventional transducers, including reduced memory consumption and accelerated computation through the reduction of time resolution in the encoded hidden state.
Through our experimentation on the AISHELL-1 dataset, we have observed that the incorporation of the time-sparse mechanism in the prediction phase leads to a notable decrease in the character error rate. This finding underscores the efficacy of the time-sparse transducer in enhancing recognition accuracy and reducing inference time consumption when employed in conjunction with various decoders.
While this study primarily focused on the development and evaluation of the time-sparse transducer, there are avenues for future research that warrant exploration. Specifically, we intend to investigate optimizing strategies for weighted average coefficients on diverse speech corpora. Such exploration will enable us to better understand the implications and generalized ability of our proposed model across different linguistic contexts.

\bibliographystyle{splncs04}
\bibliography{mybibliography}
%
% \begin{thebibliography}{8}
% \bibitem{ref_article1}
% Author, F.: Article title. Journal \textbf{2}(5), 99--110 (2016)

% \bibitem{ref_lncs1}
% Author, F., Author, S.: Title of a proceedings paper. In: Editor,
% F., Editor, S. (eds.) CONFERENCE 2016, LNCS, vol. 9999, pp. 1--13.
% Springer, Heidelberg (2016). \doi{10.10007/1234567890}

% \bibitem{ref_book1}
% Author, F., Author, S., Author, T.: Book title. 2nd edn. Publisher,
% Location (1999)

% \bibitem{ref_proc1}
% Author, A.-B.: Contribution title. In: 9th International Proceedings
% on Proceedings, pp. 1--2. Publisher, Location (2010)

% \bibitem{ref_url1}
% LNCS Homepage, \url{http://www.springer.com/lncs}. Last accessed 4
% Oct 2017
% \end{thebibliography}
\end{document}